\documentclass[twocolumn,aps,showpacs,amssymb,floatfix]{revtex4}
\newcommand{\be}{\begin{equation}}
\newcommand{\ee}{\end{equation}}
\newcommand{\beq}{\begin{eqnarray}}
\newcommand{\eeq}{\end{eqnarray}}
\usepackage{bm}
\usepackage{graphicx}%
\usepackage{epsf}
\usepackage{epsfig}

\begin{document}
\setcounter{figure}{\arabic{figure}}

\title{The nucleon 
electromagnetic form factors from Lattice QCD}
\author{C.~Alexandrou~$^a$, G. Koutsou~$^a$
 J.~W.~Negele~$^b$ and A. Tsapalis~$^c$}
\affiliation{{$^a$ Department of Physics, University of Cyprus, CY-1678 Nicosia, Cyprus}\\
{$^b$ 
Center for Theoretical Physics, Laboratory for
 Nuclear Science and Department of Physics, Massachusetts Institute of
Technology, Cambridge, Massachusetts 02139, U.S.A.}\\
{$^c$ 
Institute for Accelerating Systems and Applications, University of Athens,
Athens,Greece.}}

\date{\today}%

\begin{abstract}
We evaluate the isovector 
nucleon electromagnetic form factors in quenched and unquenched
QCD on the lattice using Wilson fermions. In the quenched theory we use
a lattice of spatial size $3$~fm at $\beta=6.0$ enabling us to reach
low momentum transfers and a lowest pion mass of about 400~ MeV. 
In the unquenched theory
we use two degenerate flavors of  dynamical Wilson fermions on
a lattice of spatial  size $1.9$~fm at $\beta=5.6$ and lowest pion
mass of about 380 MeV enabling comparison with the results 
obtained in the quenched
theory. We find that unquenching effects are small for the pion masses
 considered in this work.  We compare our lattice results to the isovector
part of the experimentally measured form 
factors.

\end{abstract}

 
\maketitle
 
 
\section{Introduction}

The elastic nucleon electromagnetic form factors  are fundamental 
quantities
characterizing important features of  neutron and  proton structure
that include their size, charge distribution and magnetization.
 An accurate  determination 
of these quantities in lattice QCD is timely and important
because of a new generation of precise experiments.
In particular, polarization experiments~\cite{Jeff} 
 that measure directly the ratio
of the proton electric to magnetic form factor, $\mu_p G_E^p/G_M^p$,
have shown a qualitative different behavior 
than the traditional Rosenbluth separation.
 The ratio
 $\mu_p G_E^p/G_M^p$ instead of being approximately constant 
 falls off
almost linearly with  the momentum transfer squared,  $q^2$. 
This means that the electric form factor falls faster
than the magnetic.
Recent reviews on the experimental situation can
be found in Refs.~\cite{Gao,Jager}.
Precise lattice data for the nucleon form factors for large values of $q^2$
will enable comparison with experiment 
and could lead to an understanding of the approach to asymptotic scaling.
Furthermore, access to 
  low momentum transfers will enable
a better determination of phenomenologically 
interesting quantities such as the root mean squared (r.m.s.) radius of the
transverse quark distribution
in the nucleon~\cite{Burkardt}.
To access small momentum transfers we need a lattice with large
spatial extent, $L$, since the smallest available momentum is
$2\pi/L$.
 Although large momentum 
transfers are in principle available on typical  lattices,  
the Fourier transform of two and three point functions 
becomes noise-dominated for momentum transfers
beyond about 2~GeV$^2$, limiting the range of high $q^2$ values that
can be extracted accurately.

In this work we 
calculate the isovector nucleon  form factors 
as a function of the momentum transfer
 in lattice QCD both  in the quenched approximation and with two dynamical 
Wilson fermions.
A recent study of the nucleon form factors was carried out in the quenched
theory using improved Wilson fermions~\cite{QCDSF}. 
The current work builds on the ingredients of the previous
lattice calculation and obtains results with higher accuracy 
 at lower momentum 
transfers and pion masses. 
This enables us to determine the momentum
dependence of the form factors  
accurately enough to have a meaningful 
comparison with experiment. A number
of phenomenologically interesting
quantities such as the r.m.s radii and dipole masses are extracted. 
Furthermore, we improve the accuracy
of the results
 by constructing an optimal
source for the nucleon allowing the maximum number of lattice momentum vectors
to contribute. The two form factors are then extracted using an overconstrained
analysis that includes all possible lattice measurements for a given value
of $q^2$.  
 For the quenched calculation we use  a lattice of size
$32^3\times 64$ at $\beta=6.0$, which corresponds to a 
lattice spacing $a\sim 0.09$~fm, obtained either by using the 
nucleon mass at the chiral limit or the Sommer scale.
In order to assess quenching effects, we also evaluate these form factors in
the unquenched theory using dynamical Wilson configurations simulated
for quark masses that give pions of mass 
690 MeV and 509 MeV on a lattice of size 
$24^3\times 40$~\cite{newSESAM} and 380 MeV on a lattice
of size $24^3\times 32$~\cite{Carsten} at $\beta=5.6$. 
The lattice spacing is about $0.08$~fm determined from the nucleon mass 
at the chiral limit.
This value is consistent with the value extracted using the
Sommer scale over the range of  quark
masses used in this work.

In lattice QCD, elastic matrix elements 
involving one-photon exchange require the evaluation of
three-point functions. The standard procedure to evaluate 
three-point functions like the one we need here,namely
$G^{N j^\mu N} (t_2, t_1 ; {\bf p}^{\;\prime}, {\bf p};\Gamma ) $,
is  to
compute the sequential propagator. This  can be done in two ways:
 In an early pioneering work,
where  matrix elements of a number of different hadronic states 
were evaluated~\cite{Leinweber},
the method of choice was to couple the photon  to a
quark at a fixed time $t_1$  carrying a fixed
momentum ${\bf q}$. Within this scheme 
the form factors can only be evaluated at one
value of the  momentum transfer. 
Since the current must have a fixed direction and a
fixed momentum, this approach is referred to as the fixed current approach. 
This method allows
one to use any initial and final state without requiring further inversions, 
which are
the time consuming part of the evaluation of three-point functions. 
In the second approach, which is the method used in recent 
studies~\cite{QCDSF, MIT, ntodelta}, one requires
 that the initial state,  created at time zero,
and  the final state,  annihilated at a later fixed time $t_2$,  have the
nucleon quantum numbers.
The current  
can couple to any intermediate time slice $t_1$ carrying any possible value
of  the lattice momentum and having any direction. Therefore, within this
scheme, with a single sequential propagator, one is able to evaluate
all possible momentum transfers and current orientations.
Since the quantum numbers of the final
state are fixed, we refer to the second method as  
 the fixed sink method.  Clearly the fixed sink method is
superior if our goal is the accurate determination of the momentum dependence 
of the nucleon form factors.

\section{Lattice techniques}
The nucleon electromagnetic matrix element 
for real or
virtual photons can be written in the form
\beq
 \langle \; N (p',s') \; | j_\mu | \; N (p,s) \rangle &=& \nonumber \\  
&\>& \hspace*{-4cm}  \biggl(\frac{ M_N^2}{E_{N}({\bf p}^\prime)\;E_N({\bf p})}\biggr)^{1/2} 
  \bar{u} (p',s') {\cal O}_{\mu} u(p,s) \; ,
\label{NjN}
\eeq
where $p(s)$ and $p'(s')$ denote initial and final momenta (spins) and 
$ M_N$ is the nucleon mass.
The operator 
${\cal O}^{\mu}$  can be decomposed in terms of the Dirac form factors
as
\be
{\cal O}_{\mu} = \gamma_\mu F_1(q^2) 
+  \frac{i\sigma_{\mu\nu}q^\nu}{2M_N} F_2(q^2) \; ,
\label{Dirac ff}
\ee
where $F_1(0)=1$ for the proton since we have a conserved current and
$F_2(0)$ measures the anomalous magnetic moment. They are connected to the  
electric, $G_E$, and magnetic, $G_M$, Sachs form factors by the relations
\beq
G_E(q^2)&=& F_1(q^2) + \frac{q^2}{(2M_N)^2} F_2(q^2)\nonumber \\
G_M(q^2)&=& F_1(q^2) + F_2(q^2) \quad .
\label{Sachs ff}
\eeq
To extract the nucleon matrix element 
from lattice measurements, we calculate, besides the three 
point function 
$G^{Nj^\mu N} (t_2, t_1 ; {\bf p}^{\;\prime}, {\bf p};\Gamma )$,
the nucleon  two-point function, $ G^{NN}(t,{\bf p})$, 
and look for a plateau in  
the large Euclidean
time behavior of the ratio 
\small
\beq
R (t_2, t_1; {\bf p}^{\; \prime}, {\bf p}\; ; \Gamma ; \mu) &=&
\frac{\langle G^{Nj^\mu N}(t_2, t_1 ; {\bf p}^{\;\prime}, {\bf p};\Gamma ) \rangle \;}{\langle G^{N N}(t_2, {\bf p}^{\;\prime};\Gamma_4 ) \rangle \;} \> \nonumber \\
&\>& \hspace*{-3.8cm}\biggl [ \frac{ \langle G^{N N}(t_2-t_1, {\bf p};\Gamma_4 ) \rangle \;\langle 
G^{NN} (t_1, {\bf p}^{\;\prime};\Gamma_4 ) \rangle \;\langle 
G^{N N} (t_2, {\bf p}^{\;\prime};\Gamma_4 ) \rangle \;}
{\langle G^{N N} (t_2-t_1, {\bf p}^{\;\prime};\Gamma_4 ) \rangle \;\langle 
G^{N N} (t_1, {\bf p};\Gamma_4 ) \rangle \;\langle 
G^{N N} (t_2, {\bf p};\Gamma_4 ) \rangle \;} \biggr ]^{1/2} \nonumber \\
&\;&\hspace*{-1cm}\stackrel{t_2 -t_1 \gg 1, t_1 \gg 1}{\Rightarrow}
\Pi({\bf p}^{\; \prime}, {\bf p}\; ; \Gamma ; \mu) \; .
\label{R-ratio}
\eeq
\normalsize
We use the lattice conserved   electromagnetic current,   $j^\mu (x)$,
symmetrized on site $x$ by taking
\be
j^\mu (x) \rightarrow \left[ j^\mu (x) + j^\mu (x - \hat \mu) \right]/ 2
\label{lattice current}
\ee
and projection matrices for the Dirac indices
\be
\Gamma_i = \frac{1}{2}
\left(\begin{array}{cc} \sigma_i & 0 \\ 0 & 0 \end{array}
\right) \;\;, \;\;\;\;
\Gamma_4 = \frac{1}{2}
\left(\begin{array}{cc} I & 0 \\ 0 & 0 \end{array}
\right) \;\; .
\ee
Throughout this work we use kinematics where the final nucleon state
 is produced at rest
and therefore ${\bf q}={\bf p}^{\prime}-{\bf p}=-{\bf p}$.
Since we aim at obtaining the full
$q^2$ dependence of the form factors, we evaluate the three point
functions with sequential inversions through the sink.
We fix $t_2=11 (12)$ in lattice units for the quenched (unquenched) Wilson
lattices and search for a plateau of 
 $ R(t_2,t_1;{\bf p}^{\; \prime}, {\bf p}\; ; \Gamma ;\mu)$ as a function of
 $t_1$. 
$Q^2=-q^2$ denotes the Euclidean momentum transfer squared.

We can extract the two Sachs form factors from the ratio of Eq.~(\ref{R-ratio}) by choosing
appropriate combinations of the direction $\mu$ of the electromagnetic current
  and projection matrices $\Gamma$.
Provided the Euclidean times $t_1$ and $t_2 - t_1$ are large
enough to filter the nucleon ground state, the ratio
becomes time independent.
 Inclusion of
hadronic states in the two- and three-point functions
leads to the expressions written in Euclidean space   

\be \Pi ( {\bf 0}, -{\bf q}\; ; \Gamma_k \; ; \mu=i)  = C \frac{1}{2
M_N} \epsilon_{ijk} \; q_j \; G_M (Q^2) 
\label{GM}
\ee

\be \Pi ( {\bf 0}, -{\bf q}\; ; \Gamma_4 \; ; \mu=i )  = C
\frac{q_i}{2 M_N} \; G_E (Q^2) 
\label{GE123}
\ee

\be \Pi ( {\bf 0}, -{\bf q}\; ; \Gamma_4 \; ; \mu = 4)  = C
\frac{E_N +M_N}{2 M_N} \; G_E (Q^2)  \; ,
\label{GE4}
\ee
where $C=
\sqrt{\frac{2 M_N^2}{E_N(E_N + M_N)}}$ is a factor due to the
normalization of the lattice states. 
The first observation regarding these expressions is that the
polarized matrix element given in Eq.~(\ref{GM}), from which
 the magnetic form factor is determined,
does not contribute for all momenta ${\bf q}$. 
In the lattice study of
the $\gamma \,N \rightarrow \Delta$ transition ~\cite{ntodelta,ntodelta0},
we dealt with a similar situation where  the naive $\Delta$ source
was not optimal in the sense that, with one sequential propagator,
 not all lattice momentum vectors resulting in the same value of $Q^2$
 contributed
and an optimal source for the $\Delta$ was needed. Similarly, here
one can construct
an optimal linear
combination for the nucleon sink that leads to
\beq  S_m({\bf q}; i)= \sum_{k=1}^3\Pi (-{\bf
q}\; ;
\Gamma_k ;\mu=i) =  \frac{C}{2M_N} \biggl\{ (p_2-p_3)\delta_{1,i} \nonumber \hspace{-0.5cm} \\
 + (p_3-p_1)\delta_{2,i} + (p_1-p_2)\delta_{3,i} \biggr\}
G_M(Q^2) 
\label{GM optimal}
\eeq 
and provides the maximal set of
lattice measurements from which  $G_M$
can be extracted requiring one sequential inversion. No such
improvement is necessary for the unpolarized matrix elements
given in Eqs.~(\ref{GE123}) and (\ref{GE4}), 
which yield $G_E$ with an additional sequential inversion.

Unlike the $\gamma \,N \rightarrow \Delta$ transition, the $\gamma
\,N \rightarrow N$  transition contains isoscalar photon
contributions. This means that disconnected loop diagrams also
contribute. These are generally difficult to
evaluate accurately since the all-to-all quark
propagator is required. In order to avoid
disconnected diagrams, we calculate the isovector form factors.
 Assuming $SU(2)$ isospin
symmetry, it follows that 
\small 
\beq 
\langle \; p \,| (
\frac{2}{3}\bar{u} \gamma^{\mu}u - \frac{1}{3}\bar{d} \gamma^{\mu}
d ) | p \rangle  - \langle \; n | ( \frac{2}{3}\bar{u}
\gamma^{\mu}u - \frac{1}{3}\bar{d} \gamma^{\mu} d ) | n \rangle \;
\nonumber \hspace{-0.9cm}\\ = \langle \; p \, | ( \bar{u}
\gamma^{\mu} u - \bar{d} \gamma^{\mu} d ) | p \rangle . 
\eeq
\normalsize 
One can therefore calculate directly the three-point
function related to the right hand side of the above relation which
provides the {\it isovector } nucleon form
factors 
\beq 
G_E (q^2) &=& G^p_E (q^2)\, - G^n_E
(q^2) , \nonumber \\
 G_M (q^2) &=& G^p_M(q^2)- G^n_M (q^2) .
\label{isovector}
 \eeq
 The isovector electric form factor,$G_E$,
 can therefore be obtained from the connected
diagram considering either the spatial
components of the electromagnetic current as given in
Eq.~(\ref{GE123}) or
 the temporal component given in Eq.~(\ref{GE4}),
 while  Eq.~(\ref{GM optimal}) 
is used for the extraction
of the isovector magnetic form factor, $G_M$.

Besides using an optimal nucleon source, the other
important ingredient in the extraction of the form factors
 is to take into
account simultaneously
in our analysis  all the lattice momentum vectors that contribute to a given 
$Q^2$. This is done by solving the overcomplete set of equations
\be
P({\bf q};\mu)= D({\bf q};\mu)\cdot F(Q^2) 
\label{overcomplete}
\ee
where $P({\bf q};\mu)$ are the lattice measurements of the ratio
given in Eq.~(\ref{R-ratio}) having statistical errors
$w_k$ and using the different sink types,
$F =  \left(\begin{array}{c}  G_{E} \\
                                    G_M \end{array}\right)$
and $D$ is an $M\times 2$ matrix which depends on 
kinematical factors with $M$ being the number of current 
directions and momentum vectors contributing to 
a given $Q^2$. We extract the form factors by 
minimizing 
\be
\chi^2=\sum_{k=1}^{N} \Biggl(\frac{\sum_{j=1}^3 D_{kj}F_j-P_k}{w_k}\Biggr)^2
\ee
using the singular value decomposition of $D$.
Given the fact
that one can have a few hundred lattice momentum vectors contributing in the evaluation
of the form factors,  the statistical precision is highly improved. 
 Phenomenologically interesting quantities
like the r.m.s. radii and magnetic moments can thus be obtained with increased
precision.
There is an additional advantage by including momentum transfers 
${\bf q}$ as well as $-{\bf q}$ in our analysis.
 The lattice conserved current given in 
Eq.~(\ref{lattice current}) differs from the local electromagnetic 
current $\bar{\psi}(x) \gamma_\mu \psi(x)$ by terms of ${\cal O}(a)$. However
when we average over ${\bf q}$ and $-{\bf q}$ these   ${\cal O}(a)$ terms
vanish.

\begin{figure}[h]
\epsfxsize=8.5truecm
\epsfysize=10.5truecm
\mbox{\epsfbox{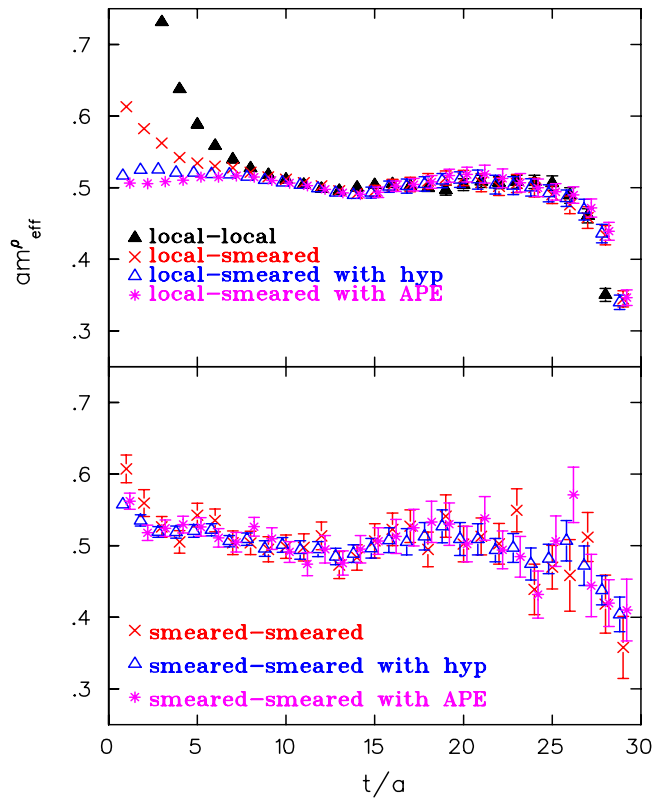}}
\vspace*{0.5cm}
\caption{The rho effective mass as a function of the time
separation on a $16^3\times 32$ quenched lattice at $\beta=6.0$ and
 $\kappa=0.153$ using Dirichlet boundary conditions in the
temporal direction. In the upper graph,
 filled triangles show results obtained with local
source and sink, crosses with Wuppertal
smeared source and local sink and open triangles (asterisks) with
Wuppertal smeared source using hypercubic (APE)  smearing for the 
gauge links used in the construction of
the hopping matrix $H({\bf x},{\bf z};U(t))$.
 The lower graph shows with crosses
results obtained using Wuppertal
smeared source and sink and with open triangles (asterisks) results with 
Wuppertal smeared source and sink where hypercubic (APE) smearing
is applied to the spatial links entering the hopping matrix. }
\label{fig:meff rho}
\end{figure}

\begin{figure}[h]
\epsfxsize=8.5truecm
\epsfysize=10.5truecm
\mbox{\epsfbox{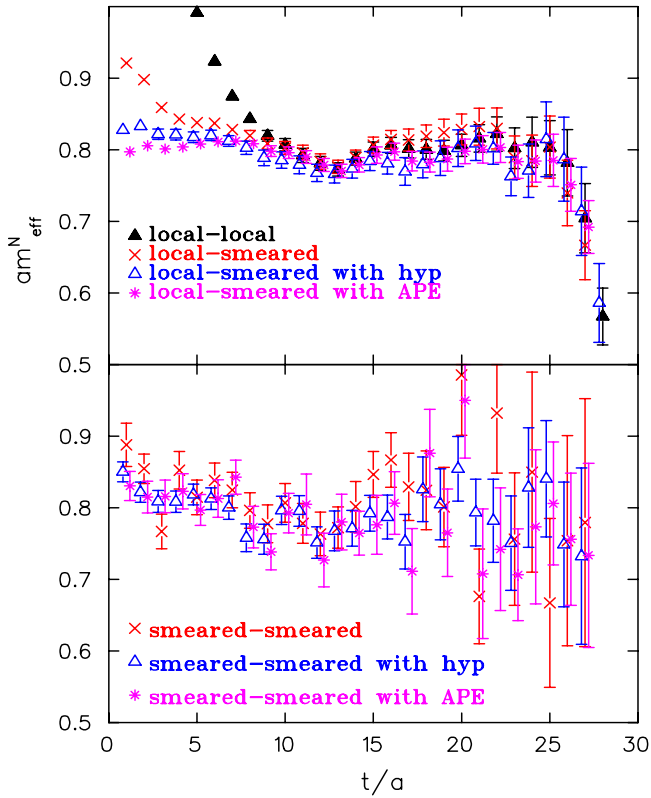}}
\vspace*{0.5cm}
\caption{The nucleon effective mass as a function of the time
separation on a $16^3\times 32$ quenched lattice at $\beta=6.0$ and
 $\kappa=0.153$. The notation is the same as that of Fig.~\ref{fig:meff rho}.}
\label{fig:meff nucl}
\end{figure}

Smearing techniques are routinely used
for achieving   ground state dominance
before the signal from the time correlators
is lost in the noisy large time limit.
We use  gauge invariant Wuppertal smearing,
$d(x,t) \rightarrow d^{\rm smear}(x,t)$,
 at the source and the sink.
We smear  the fermion interpolating fields according 
to~\cite{Guesken-Alexandrou}
\be
d^{\rm smear}({\bf x},t) = \sum_{\bf z} F({\bf x},{\bf z};U(t)) d({\bf z},t)
\ee
with the gauge invariant smearing function constructed from the
hopping matrix $H$:
\be
F({\bf x},{\bf z};U(t)) = (1+\alpha H)^n({\bf x},{\bf z};U(t)),
\ee
where 
\be
H({\bf x},{\bf z};U(t))= \sum_{i=1}^3 \biggl( U_i({\bf x},t)\delta_{{\bf x,y}-i} +  U_i^\dagger({\bf x}-i,t)\delta_{{\bf x,y}+i}\biggr).
\ee
It is well known that smearing introduces gauge noise increasing the
errors on the extracted effective masses in particular when Wuppertal smearing
is applied to both source and sink. An efficient way to reduce the
the ultraviolet fluctuations is to smooth the gauge fields 
at the time slice of the source or the sink
where Wuppertal smearing is carried out.
One can apply various smoothing techniques such as APE~\cite{APE}, 
stout~\cite{stout} or
hypercubic~\cite{HYP} smearing on the
gauge fields that are used in the hopping matrix. We found that both
APE and hypercubic smearing reduce the noise and at the same time improve 
further the ground state overlap. In Figs.~\ref{fig:meff rho} and 
\ref{fig:meff nucl} we
show the effective mass for the rho meson and the nucleon respectively, using
a lattice of size $16^3\times 32$ with Dirichlet boundary conditions in 
the temporal direction to utilize the full time extent of the lattice.
When only the source is smeared, both APE and  hypercubic smearing improve 
the ground state overlap to such an extent that the plateau value
is reached within a time separation as short as two time slices.
When we apply Wuppertal smearing both to the source and sink,
  we see a reduction in the 
gauge noise in particular when hypercubic smearing
is used. There is little effect
on improving ground state dominance since Wuppertal smearing on both 
source and sink very effectively cuts down
excited state contributions already after
 a time separation of a couple of time  slices.
 Given the better noise reduction observed when we use hypercubic
smearing  
we choose to apply this smearing  to the links that enter the
hopping matrix $H$. For the parameters that enter
the hypercubic smearing we  use the same
ones as those of Ref.~\cite{HYP}. The parameters 
for the Wuppertal smearing 
are then optimized so that ground state dominance for the nucleon is
optimal. We find that the parameters $\alpha=4$ and $n=50$ produce optimal 
results. These values are the same as those obtained without
applying hypercubic
smearing.
Whereas Wuppertal smearing is applied to the source and the sink
 in all our computations to ensure ground state
dominance at the time slice of the insertion of the electromagnetic current,
hypercubic smearing is only done in the case of the unquenched
configurations. This is because  self averaging 
is less effective on smaller lattices 
causing the gauge noise to be more
severe in the unquenched case where
the simulations were done on a smaller lattice.

\section{Discussion of experimental results}

As explained in section II, we only compute the isovector part of the form
factors given in Eq.~(\ref{isovector}). Therefore,
to compare with experiment, it is necessary to extract from the
experimentally available proton and neutron data the isovector contribution.
In order to do this we need to interpolate the proton and neutron data 
to the same $Q^2$ values. 
\begin{figure}[h]
\epsfxsize=8.5truecm
\epsfysize=5.5truecm
\mbox{\epsfbox{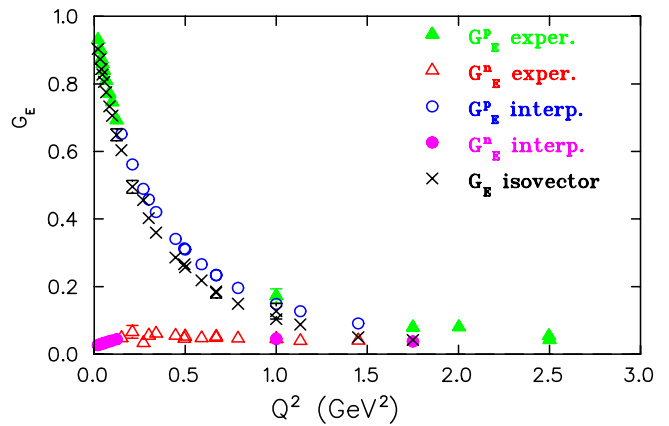}}
\caption{The isovector electric form factor, $G_E$,
 extracted by interpolation from 
the measured proton and neutron electric form factors.}
\label{fig:GE experiment}
\end{figure}

\begin{figure}[h]
\epsfxsize=8.5truecm
\epsfysize=5.5truecm
\mbox{\epsfbox{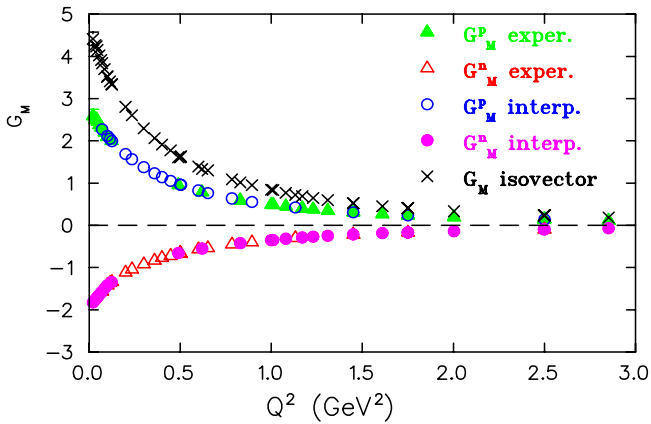}}
\caption{The isovector magnetic form factor, $G_M$,
 extracted by interpolation from 
the measured proton and neutron magnetic form factors.}
\label{fig:GM experiment}
\end{figure}

\begin{figure}[h]
\epsfxsize=8.5truecm
\epsfysize=5.5truecm
\mbox{\epsfbox{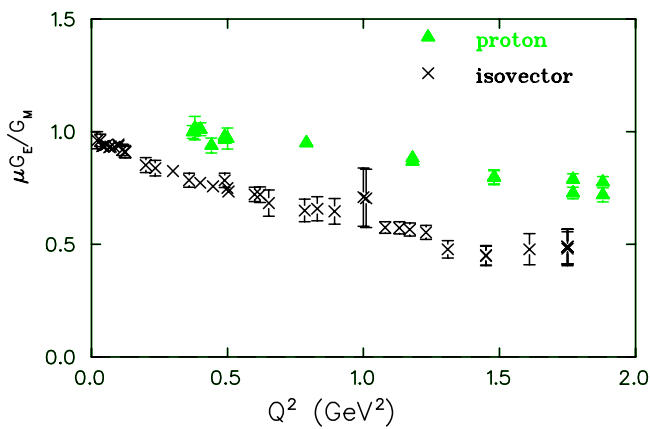}}
\caption{The ratio of isovector form factors $G_E$ over
$G_M$ as compared to the corresponding ratio of proton
form factors from recent polarization experiments~\cite{Jeff}.}
\label{fig:GEoverGM experiment}
\end{figure}
In Fig.~\ref{fig:GE experiment} we show the
proton and neutron data for the electric 
form factor~\cite{exp_proton,exp_neutron}. 
As can
be seen, we need to extrapolate the neutron electric form factor $G^n_E$
at low momentum transfers and the proton electric form factor $G^p_E$
at  intermediate $Q^2$ values in the range $0.25 <Q^2<1.0$~GeV.
 In order to interpolate
the neutron data 
we consider the Galster parametrization~\cite{Galster}
\be
G^N_E(Q^2) = \frac{-\mu_n\tau}{1+5.6\tau} G_d(Q^2)  \; ,
\label{galster}
\ee
where $\tau=Q^2/4{M_N}^2$, $G_d(q^2)=1/(1+Q^2/0.71)^2$ and $\mu_n=-1.91315$,
which  provides a good description of the data. We 
calculate the derivative needed
for the interpolation between measured data 
using the Galster parametrization. Similarly, in order to interpolate the 
proton data,
we fit to a dipole form and then use the fitted form
 to compute the derivative needed for the
interpolation. Having $G^n_E$ and $G^p_E$ at the same value of $Q^2$, we then
find the isovector contribution via Eq.~(\ref{isovector}) and
 plot the resulting
isovector $G_E$ in Fig.~\ref{fig:GE experiment}. As expected, the difference
between $G^p_E$ and the isovector part is small due to the smallness of
$G^n_E$. A similar analysis is done for the magnetic form factor using
the data of Ref.~\cite{exp_proton,exp_neutron}. 
The derivative needed for the interpolation is computed either from the best 
dipole fit function to the available data or, in the cases
where we have two measurements
close to the value of $Q^2$ that we are interested in, by using a finite
difference approximation to the derivative. 
The resulting isovector magnetic form factor is  
shown in Fig.~\ref{fig:GM experiment}.
In Fig.~\ref{fig:GEoverGM experiment} we compare the extracted isovector 
ratio  $\mu G_E/G_M$ to recent measurements of the proton ratio 
$\mu_p G^p_E/G^p_M$
that showed an unexpected $Q^2$ dependence~\cite{Jeff,exp_ratio}. 
The isovector
ratio for $Q^2 < 1$~GeV$^2$ decreases faster with $Q^2$ 
than $\mu_p G^p_E/G^p_M$, whereas 
for 
$Q^2 > 1$~GeV$^2$ it remains approximately constant. 
One of the goals is to compare this behavior with lattice calculations.

\section{Lattice results}
As pointed out in the Introduction, the purpose of this work is to obtain
accurate results over a large range of momentum transfers. 
For this reason 
the quenched calculation is done on a lattice of size  $32^3\times 64$ 
enabling us to reach momentum transfers as low as about 0.15~GeV$^2$.
The highest momentum transfer that is accessible at $\beta=6.0$
is $2\pi/a\sim 13$~GeV.
However statistical errors do not allow us to reach this maximum value. 
For the unquenched calculation we
use configurations generated
by the SESAM collaboration~\cite{newSESAM} on a lattice of size 
$24^3\times 40$  and the DESY-Zeuthen group~\cite{Carsten} 
on a lattice of size $24^3\times 32$ at $\beta=5.6$. At this value of 
$\beta$ the lattice spacing is close  enough to the 
lattice spacing of the quenched lattice so
that finite $a$-effects are comparable. Differences between
the two evaluations can then 
be attributed to unquenching effects. 
In Table~\ref{table:parameters} we give the 
parameters of our calculation.

\begin{table}[h]
\caption{In the first column we give the number of configurations
and  in the second column the value of the hopping parameter, $\kappa$, 
that fixes the bare quark mass.
 In the third and fourth columns we give
 the pion and nucleon mass in lattice
units. The values of the lattice spacing $a$ are determined from the
mass of the nucleon at the chiral limit. The unquenched configurations
at $\kappa=0.1575$ and $\kappa=0.1580$ are provided by the SESAM 
collaboration~\cite{newSESAM} and at $\kappa=0.15825$ by 
the DESY-Zeuthen group~\cite{Carsten}.} 
\vspace*{.2cm}
\label{table:parameters}
\begin{tabular}{|c|c|c|c|}
\hline
\multicolumn{1}{|c|}{number of confs } &
\multicolumn{1}{ c|}{$\kappa$ } &
\multicolumn{1}{ c|}{$am_\pi$ } &
\multicolumn{1}{ c|}{$aM_N$ } 
\\
\hline
\multicolumn{4}{|c|}{Quenched $32^3\times 64$ \hspace*{0.5cm} $a^{-1}=2.14(6)$~GeV}
 \\ \hline
  200            &  0.1554 &  0.263(2) & 0.592(5)     \\
  200            &  0.1558 & 0.229(2) &  0.556(6) \\
  200            &  0.1562 & 0.192(2) &  0.518(6)  \\
    &  $\kappa_c=$0.1571   & 0.       &  0.439(4) \\
\hline
\multicolumn{4}{|c|}{Unquenched $24^3\times 40$  \hspace*{0.5cm}
$a^{-1}=2.56(10)$ GeV} 
 \\\hline
 185                &  0.1575  & 0.270(3) & 0.580(7) \\
 157                &  0.1580  & 0.199(3) & 0.500(10)  \\
\hline
\multicolumn{4}{|c|}{Unquenched $24^3\times 32$  \hspace*{0.5cm}
$a^{-1}=2.56(10)$ GeV} 
\\\hline
 200                &  0.15825 & 0.150(3) & 0.423(7)    \\
                    & $\kappa_c=0.1585$& 0. & 0.366(13)\\
\hline
\end{tabular}
\end{table} 

\begin{figure}[h]
\epsfxsize=8.5truecm
\epsfysize=10.5truecm
\mbox{\epsfbox{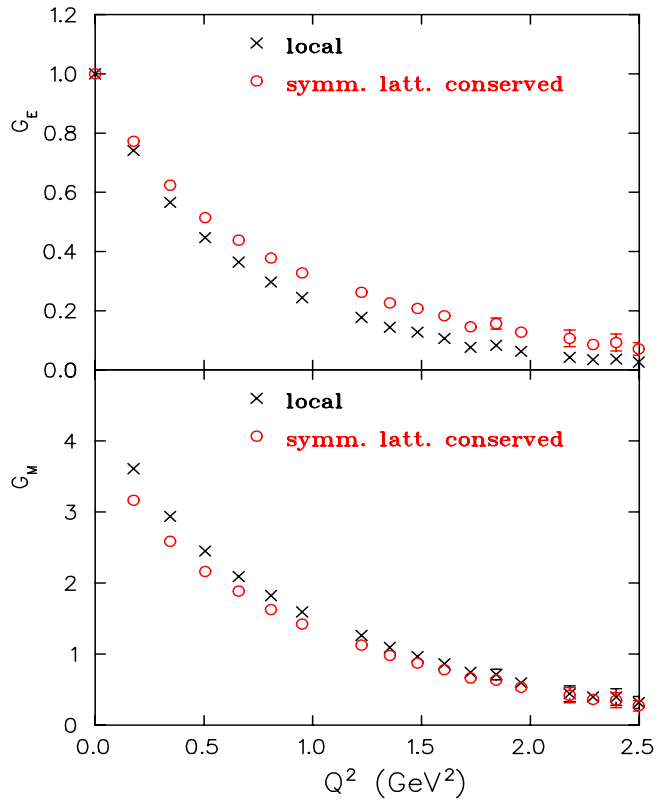}}
\caption{$G_E$ and $G_M$  as functions of $Q^2$ in the quenched theory at
 $\kappa=0.1554$ extracted
using the local current $\bar{\psi}(x)\gamma_\mu \psi(x)$
 (open circles) and the symmetrized lattice conserved
 electromagnetic current (crosses).}
\label{fig:current compare}
\end{figure}

\begin{figure}[h]
\epsfxsize=8.5truecm
\epsfysize=10.5truecm
\mbox{\epsfbox{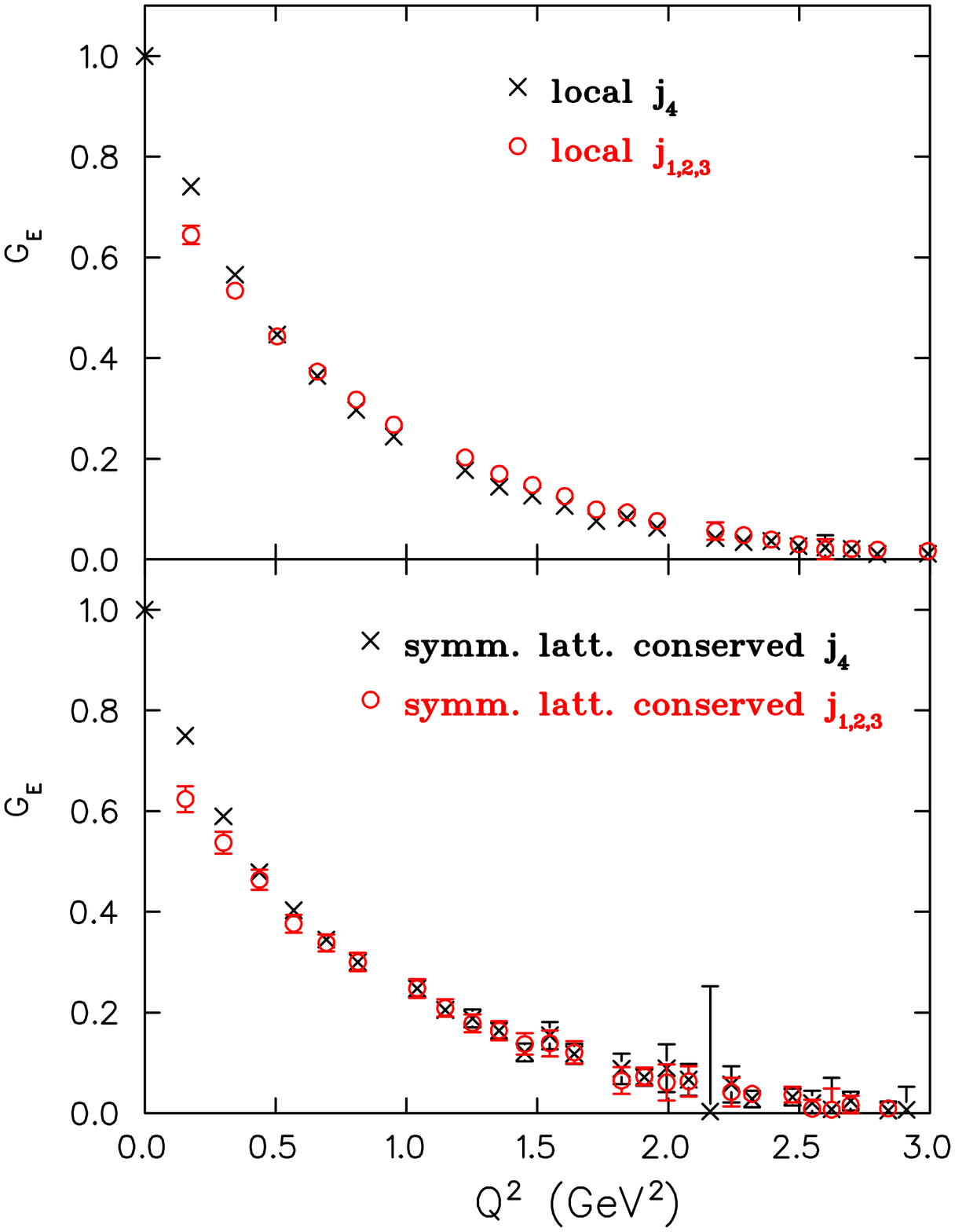}}
\caption{$G_E$ as a function of $Q^2$ in the quenched theory at
 $\kappa=0.1554$ extracted
using Eq.~(\ref{GE123}) (open circles) when the 
electromagnetic current is
in the spatial  direction and using Eq.~(\ref{GE4}) (crosses)
with the electromagnetic
current in the temporal direction. The top graph show results
extracted using the local current $\bar{\psi}(x)\gamma_\mu \psi(x)$
and the lower using the symmetrised lattice conserved current.}
\label{fig:GE compare}
\end{figure}

 The lattice spacing is determined from the mass of the nucleon in the
chiral limit. We use two different {\it Ans\"atze} for extrapolating to
the chiral limit: one is $aM_N=aM_N(0) + c_0 m_\pi^2$ and the other
$aM_N=aM_N(0) + c_1 m_\pi^2+c_2 m_\pi^3$. 
This provides an estimate for
the systematic error in the extrapolated value of the nucleon mass
 which dominates the overall error quoted in Table~\ref{table:parameters}
for the lattice spacing.
We would like to point out that using the  Sommer scale to set
the scale and taking $r_0=0.5$~fm, we find values that are consistent
with the ones extracted from the nucleon mass. For  
the quenched lattice
using for $r_0/a$ the values given in Ref.~\cite{Sommer}, we obtain
$a^{-1}=2.15$~GeV ($a=0.093$~fm). For the unquenched lattice,
 the same definition gives at $\kappa=0.1575$,
 $\kappa=0.1580$ and $\kappa=0.15825$  $a^{-1}=2.42(4)$~GeV,
$a^{-1}=2.47(3)$~GeV  and $a^{-1}=2.56(6)$~GeV~\cite{Carsten} respectively.
These values are consistent
with the value of $a^{-1}=2.56(10)$~GeV  
extracted from the nucleon mass at the chiral limit.
The choice of the lattice spacing affects the physical value of $Q^2$ but
not the values of $G_E$ and $G_M$ since they are computed
in dimensionless units.
In other words, a change in the lattice spacing $a$   
stretches the curves for $G_E$ and $G_M$ along the $Q^2$ axis and thus
changes the slope.

\begin{figure}[h]
\epsfxsize=8.5truecm
\epsfysize=5.5truecm
\mbox{\epsfbox{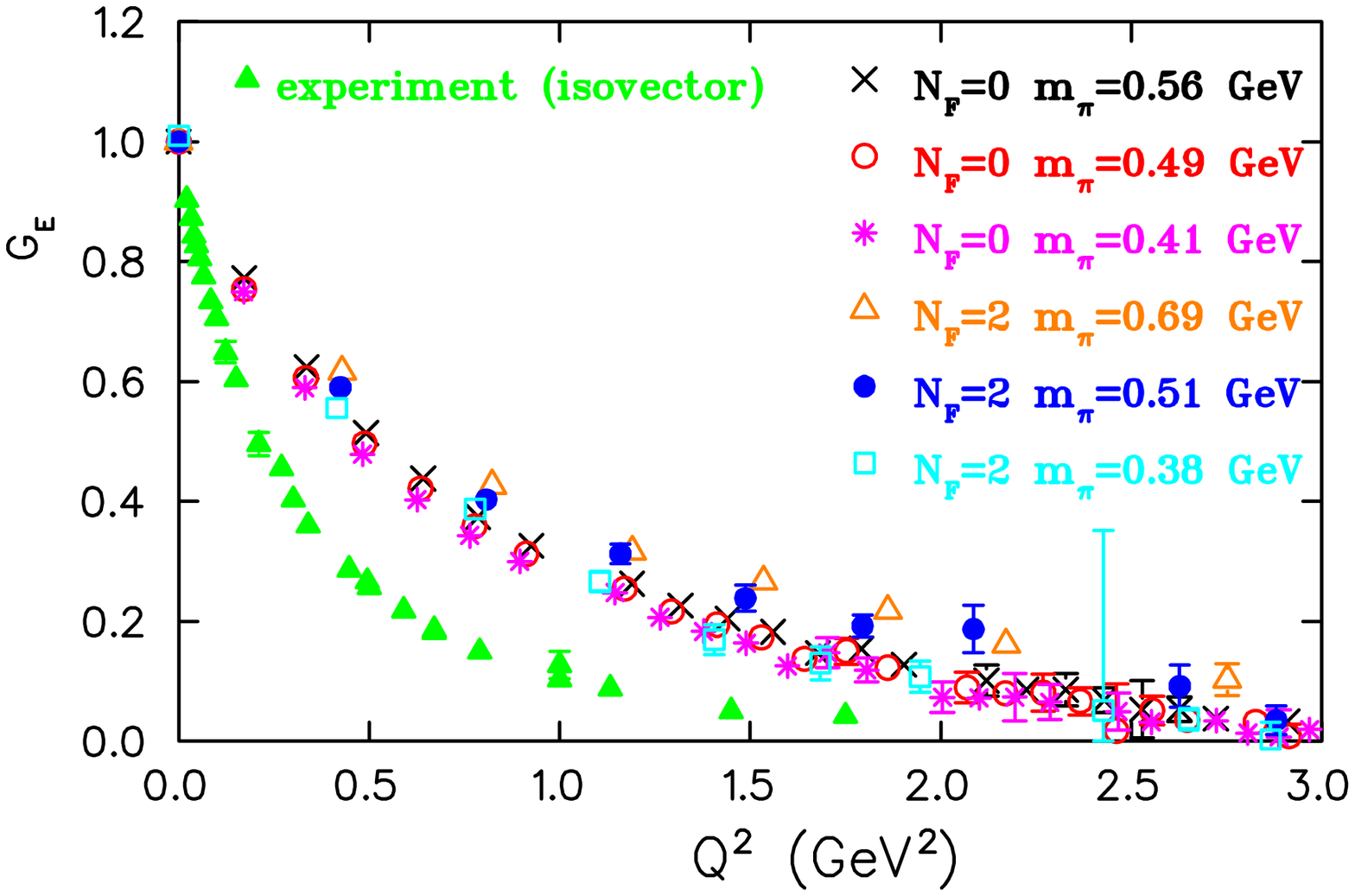}}
\caption{The isovector form factor $G_E$ as a function of $Q^2$. 
 We show quenched lattice results at
$\kappa=0.1554$ (crosses), at $\kappa=0.1558$ (open circles) 
and at $\kappa=0.1562$ (asterisks). 
The unquenched results are shown at $\kappa=0.1575$ (open triangles)
$\kappa=0.1580$ (filled circles)
and at $\kappa=0.15825$ (open squares).
The filled triangles show  
 experimental results for the isovector electric form factor
extracted using the analysis described 
in Section~III and 
data from Refs.~\cite{exp_proton,exp_neutron}.
}
\label{fig:GE}
\end{figure}

\begin{figure}[h]
\epsfxsize=8.5truecm
\epsfysize=5.5truecm
\mbox{\epsfbox{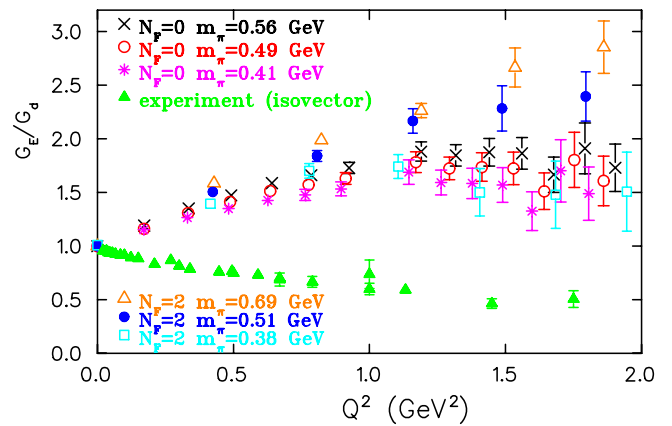}}
\caption{The isovector electric form factor, $G_E$, divided
by the proton dipole form factor as a function of $Q^2$.
The notation is the same as in Fig.~\ref{fig:GE}.}  
\label{fig:GEoverGd}
\end{figure}

\begin{figure}[h]
\epsfxsize=8.5truecm
\epsfysize=5.5truecm
\mbox{\epsfbox{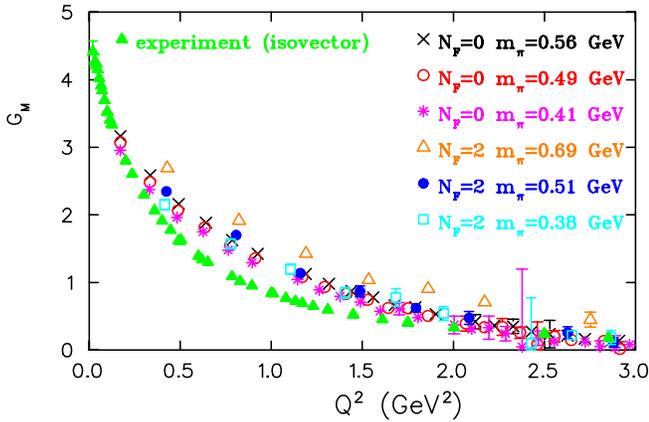}}
\caption{The isovector magnetic form factor, $G_M$, as a function of $Q^2$. 
The notation is the same as in Fig.~\ref{fig:GE}.} 
\label{fig:GM}
\end{figure}

\begin{figure}[h]
\epsfxsize=8.5truecm
\epsfysize=5.5truecm
\mbox{\epsfbox{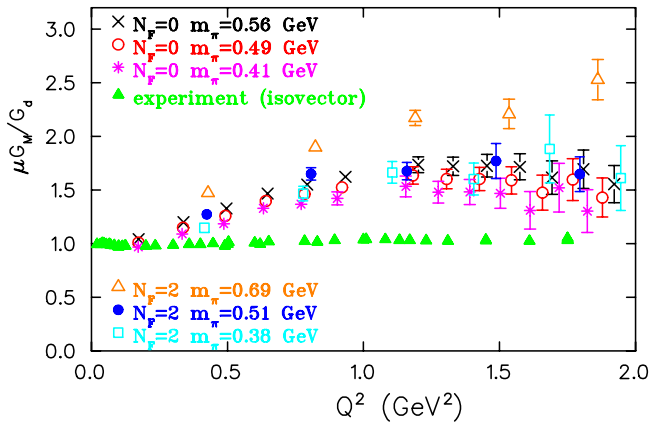}}
\caption{The isovector form factor, $G_M$, divided
by the proton dipole form factor taking $\mu=4.71$, 
as a function of $Q^2$.
The notation is the same as in Fig.~\ref{fig:GE}.}
\label{fig:GMoverGd}
\end{figure}

\begin{figure}[h]
\epsfxsize=8.5truecm
\epsfysize=10.5truecm
\mbox{\epsfbox{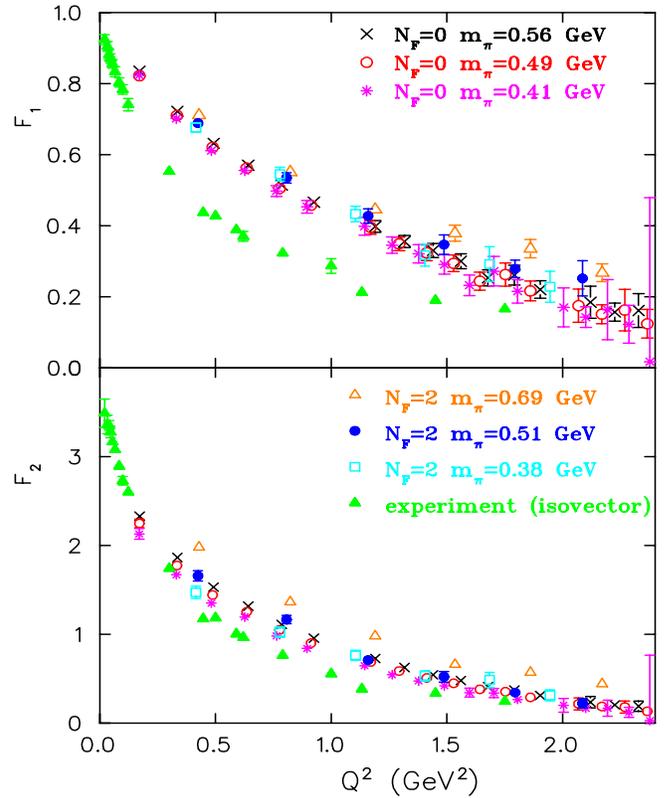}}
\caption{The isovector form factors $F_1$ (top) and $F_2$ (bottom), 
as a function of $Q^2$. 
The notation is the same as in Fig.~\ref{fig:GE}.}
\label{fig:F1 F2}
\end{figure}

\begin{figure}[h]
\epsfxsize=8.5truecm
\epsfysize=5.5truecm
\mbox{\epsfbox{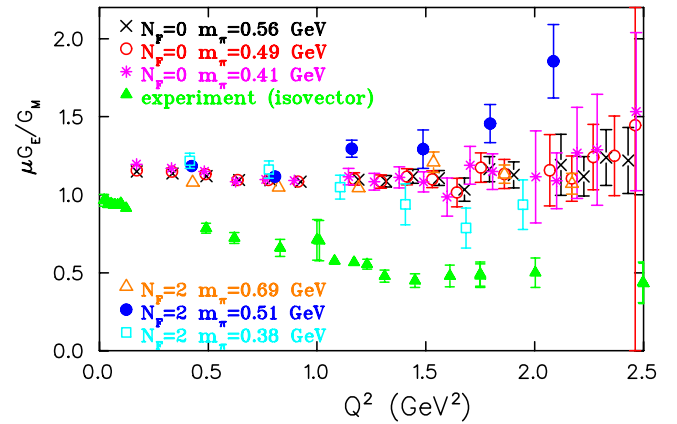}}
\caption{The isovector ratio, $\mu G_E/G_M$, as a function of $Q^2$. 
The notation is the same as in Fig.~\ref{fig:GE}.}
\label{fig:GEoverGM}
\end{figure}

\begin{figure}[h]
\epsfxsize=8.5truecm
\epsfysize=5.5truecm
\mbox{\epsfbox{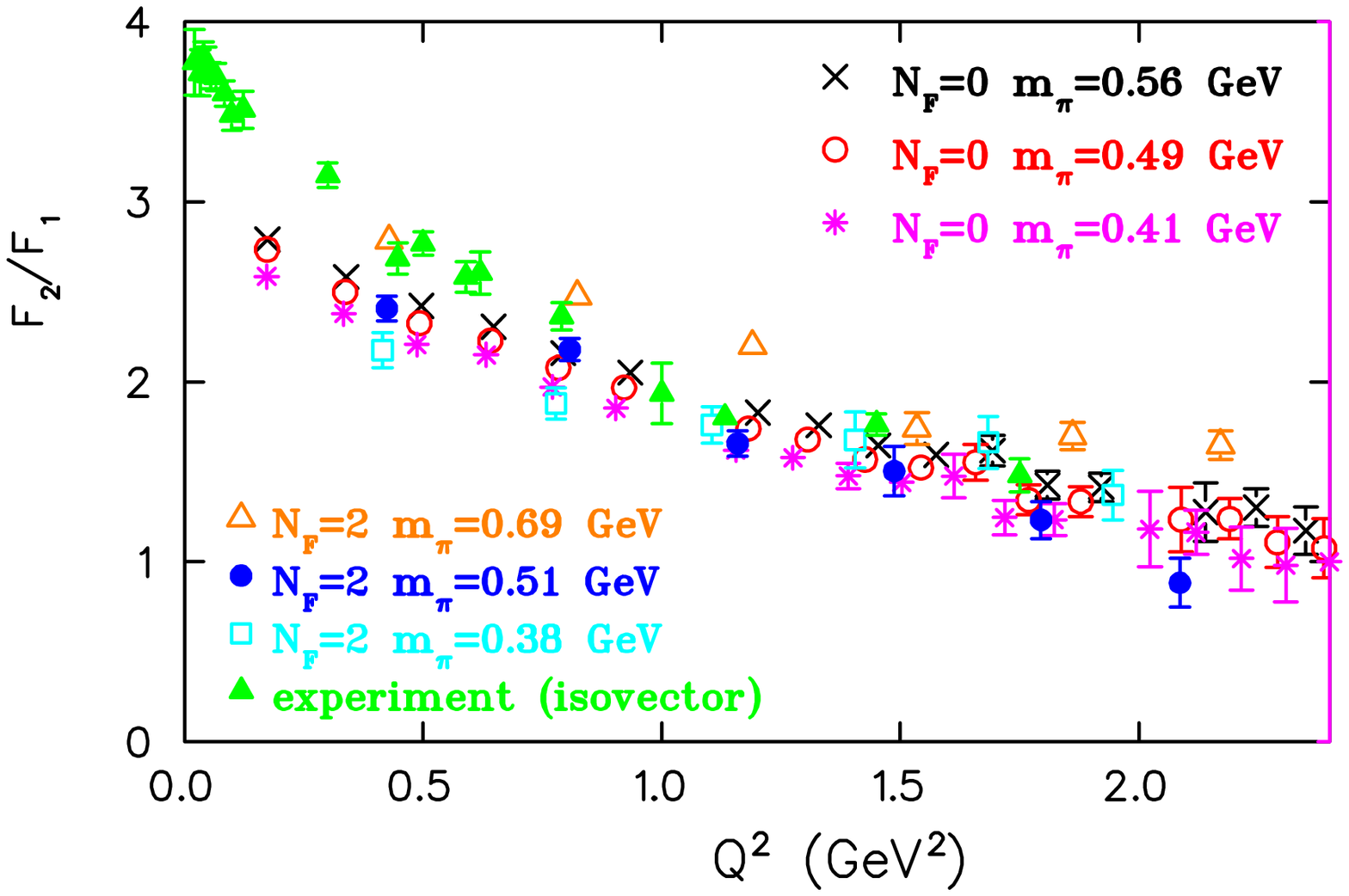}}
\caption{The isovector ratio, $F_2/F_1$, as a function of $Q^2$. 
The notation is the same as in Fig.~\ref{fig:GE}.}
\label{fig:F2overF1}
\end{figure}

All the  quenched results for the form factors are obtained 
using 200 configurations and three values of 
the  hopping parameter $\kappa$. The values
of $\kappa$ chosen are
$0.1554, 0.1558$ and 0.1562 and give a ratio of pion to rho mass
$m_\pi/m_\rho=0.64, 0.59$ and $0.50$ respectively. 
We compare to unquenched results simulated using two flavors of Wilson
fermions at $\kappa=0.1575$,
 $\kappa=0.1580$~\cite{newSESAM} and $\kappa=0.15825$~\cite{Carsten}
 that give a ratio of pion
to rho mass of $0.69$, $0.56$ and $0.45$ respectively.
The nucleon isovector elastic form factors are extracted by solving the 
overconstrained
set of equations defined in Eq.~(\ref{overcomplete}).

In Fig.~\ref{fig:current compare} we compare results for the nucleon form
factors obtained in the quenched theory at $\kappa=0.1554$ using the
local current and the lattice conserved current given in 
Eq.~(\ref{lattice current}). The local current is renormalized 
requiring charge conservation. The value of the renormalization
constant that we find is $G_E^{-1}(0)=Z_{\rm v}=0.67$.
 As we already pointed out,
averaging over all directions of ${\bf q}$, eliminates
order $a$-terms in the lattice conserved current that arise
from using fermionic fields at neighboring sites and the link 
variable that joins them. Despite the elimination of  order $a$-terms,
Fig.~\ref{fig:current compare} shows a discrepancy between results obtained
 using the local current and
the lattice conserved current. In the case of the electric 
form factor, the results from the local current have smaller values, whereas
for the magnetic case they have higher values. This means that these differences will
be amplified in the ratio of the two form factors. Since in our
approach the order $a$-terms are eliminated in the conserved
current, there is no other obvious improvement, as
far as the current is considered, that we can implement. Given that
the lattice current is the one that ensures charge conservation,
the consistent approach is to use this current.
A further argument for this choice is provided by considering the
electric form factor   
$G_{E}$, which can be evaluated using
  Eq.~(\ref{GE123}) or  Eq.~(\ref{GE4}). 
In Fig.~\ref{fig:GE compare} we compare the electric form factor extracted 
using Eq.~(\ref{GE4}) with that extracted from Eq.~(\ref{GE123}). As can be seen
for $Q^2$ larger than about $0.5$~GeV$^2$ there is perfect agreement
when using the lattice conserved current.
The disagreement at lower $Q^2$ can be understood from the dependence
on the momentum transfer  appearing in the right hand side 
of Eq.~(\ref{GE123}). As ${\bf q}\rightarrow {\bf 0}$ the right hand side
of Eq.~(\ref{GE123}) tends to zero. Inverting to obtain $G_E$ from the measured
$\Pi({\bf 0},-{\bf q};\Gamma_4;\mu=i)$ becomes inaccurate resulting in
an erroneous value for $G_E$. Using the local current on the other hand,
we observed small differences between the results 
obtained using Eq.~(\ref{GE4}) and
Eq.~(\ref{GE123}) up to values of $Q^2$ as large as 2~GeV$^2$,
which indicates a lattice artifact. Therefore, given charge
conservation and consistency of the results coming from
two different determinations of $G_E$, we conclude that
 the lattice conserved
current is the best choice for the evaluation of the form factors
within our current framework. 
Furthermore for $Q^2< 0.5$~GeV$^2$, we will only
use Eq.~(\ref{GE4}) for the determination of $G_E$ whereas for higher
$Q^2$ values both Eqs.~(\ref{GE123}) and (\ref{GE4}) will be used.

In Fig.~\ref{fig:GE} we show  the results for the electric
form factor at  three 
values of $\kappa$ for the quenched and the unquenched cases.
On the scale of this
figure, only a weak quark mass dependence is seen.
Both quenched and  unquenched results decrease as the quark mass decreases,
yielding a larger slope at small $Q^2$, which is the expected behavior. 
 In order
to better resolve differences in  our data, 
we plot in Fig.~\ref{fig:GEoverGd}
the ratio of the electric form factor
 to the proton dipole form factor, $G_E/G_d$.
 Both  quenched and unquenched results are clearly
 higher than the experimentally determined data, decreasing
with the quark mass. The unquenched results in general 
show a stronger quark mass dependence leading to 
smaller values in the chiral limit.
 The main observation, however, is that
both quenched and unquenched results have a different  $Q^2$ dependence
as compared to the  results extracted from experimental measurements:
 The lattice data have a
positive slope at small $Q^2$ whereas experiment favors  a  negative
slope. The two main uncertainties regarding the lattice results
are finite $a$-effects and whether we are close enough to the chiral limit.
Since unquenched Wilson configurations are only
available at this lattice spacing,
assessing whether finite $a$-effects can explain this behavior is beyond
the scope of the present study. Also dynamical 
Wilson configurations at smaller quark masses
on large enough volumes are not available so at present we cannot evaluate these
form factors closer to the chiral limit.

The evaluation of  the magnetic form factor $G_M$ 
 is done using  Eq.~(\ref{GM optimal}), which employs the optimal source.
The results for the magnetic form factor are shown in Fig.~\ref{fig:GM}
and, on the scale of this figure, the lattice
results are closer to 
experiment than the results for $G_E$.
 Again, quenched and unquenched results decrease with the quark mass
with the unquenched results showing a stronger quark mass dependence.
 The stronger quark mass dependence of the unquenched data
at low $Q^2$ is more clearly 
seen in Fig.~\ref{fig:GMoverGd}
where we plot the ratio $G_M/G_d$. Again in the chiral limit
we expected a reduction in the value of $G_M$  bringing lattice results
 closer to experiment. 
It is worth noting  that the experimentally
determined isovector form factor is very well described by the dipole form $G_d(Q^2)$
whereas the lattice data
 clearly   show deviations
from the dipole form at least
for the mass range considered in this work.
 To directly compare,
however, to experiment one has to carry out a chiral extrapolation of 
$G_E$ and $G_M$. This is discussed in the next section.
Ideally one must also carry out the continuum limit 
using lattices of different values of $a$, which however
is beyond the scope of the present work. 
For completeness, we  show in Fig.~\ref{fig:F1 F2}
the form factors $F_1$ and $F_2$, which are a linear combination of $G_E$
and $G_M$. What can be seen is that, in the case of $F_1$, the lattice
results show only a very weak  increase in the slope as the quark decreases. 
The slope of $F_1$ is directly related to the
transverse size of the hadron~\cite{Burkardt} and one expects an increase
in the slope as the quark mass decreases, which is not observed 
in the lattice data.
 On the other hand, in the case of $F_2$ one observes a stronger
quark mass dependence. 
This stronger quark mass
dependence potentially can lead to agreement with experiment 
 after the chiral extrapolation is carried out. 
In the case of $F_1$,
given the weaker quark mass dependence
and the larger deviation from experiment, one would require  a non-trivial
mass dependence at small quark masses to reconcile the lattice data
with experiment.
The experimentally
interesting ratio
of form factors, $\mu G_E/G_M$, is shown in Fig.~\ref{fig:GEoverGM}.
As can be seen it 
shows very little dependence on the
quark mass  and, modulo finite $a$-effects,
 it can already be compared to experiment. The ratios
obtained in the 
quenched and the unquenched theory are in agreement with each other but 
disagree with the behavior extracted from experiment.
This disagreement is also clearly seen 
in the ratio $F_2/F_1$ shown in Fig.~\ref{fig:F2overF1} especially
at small $Q^2$.

\section{Extrapolation to the chiral limit}

\begin{figure}[h]
\epsfxsize=8.truecm
\epsfysize=10truecm
\mbox{\epsfbox{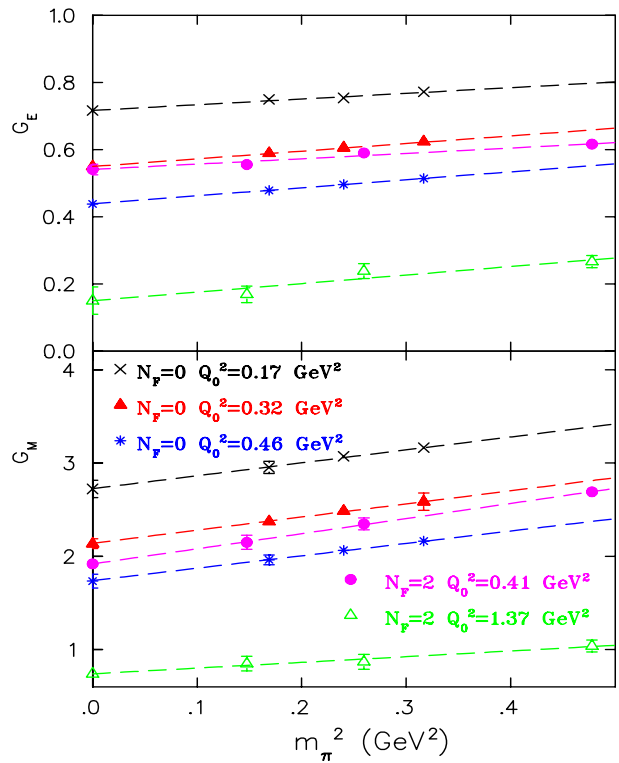}}
\caption{The isovector electric (upper) and magnetic(lower) form factor
as a function of $m_{\pi}^2$ for  the three lowest  $Q^2$ values
  available on our quenched lattice. With $Q_0^2$ we denote
 the momentum transfer square in the chiral limit.
  Unquenched results are shown with the filled circles at the
 lowest available $Q^2$-value and with the open triangles at $Q_0^2=1.37$~GeV$^2$.}
\label{fig:GM_GE_mpi2}
\end{figure}

In order to compare our results for $G_E$ and $G_M$ 
with experiment, we must extrapolate the lattice
results to the chiral limit. 
The quark masses used in this work correspond to pion masses in the 
range  560 to 410 MeV in the quenched theory and
690 to 380 MeV in the unquenched theory. Pion cloud effects are expected to be 
small in this range of pion masses and therefore we expect the 
results to show a linear dependence in the pion mass squared, $m_\pi^2$. To 
carry out correctly the chiral extrapolation of
the form factors one
would need chiral perturbation theory in the range
of pion masses that we have results and valid
 for momentum
transfers $Q^2$ up to about 2~GeV$^2$. The only chiral expansion 
for the form factors presented recently is limited to
 small momentum transfers~\cite{ff chiral}.
On the other hand, one expects that for values of 
$Q^2 \stackrel{>}{\sim}0.5 $~GeV$^2$ non-analytic terms are suppressed and a linear
dependence in $m_\pi^2$ provides a good
description to the data.    
In Fig.\ref{fig:GM_GE_mpi2}
 we plot the magnetic and electric form factors for the three lowest 
$Q^2$ values as a function of $m_\pi^2$. 
We used Eq.~(\ref{GE4}) to extract
the electric form factor since  $Q^2<0.5$~GeV$^2$
and only Eq.~(\ref{GE4}) yields reliable results.  For these lowest
values of $Q^2$ non-analytic terms could be important and should
be visible as the pion mass decreases. As can be seen in this figure, 
a linear dependence is consistent for the quenched data 
 at the three lowest $Q^2$ values.
The unquenched data are also consistent with a linear behavior. 
This
is indeed what is  observed also for the higher $Q^2$ values
 as can be seen in Fig.~\ref{fig:GM_GE_mpi2} where we show the unquenched
results at $Q_0^2=1.37$~GeV$^2$ where by $Q_0^2=2M_N(E_N-M_N)$ we denote the
 momentum transfer squared in the chiral limit obtained by using
 the physical nucleon mass.
 It is therefore reasonable to
  extrapolate the form factors linearly in $m_\pi^2$
  to obtain results in the chiral limit.
 Since  $Q^2$ depends on the mass of
 the nucleon it changes with the quark mass and we need
 to extrapolate form factors evaluated at somewhat different $Q^2$.
 To leading order $Q^2$ decreases linearly with $m_\pi^2$. Therefore
 we perform a fit to the form
 \be
 f(Q^2,m_\pi^2)=f(Q_0^2,0) + A m_\pi^2
 \label{q2}
 \ee
 where we extract the form factor at the chiral limit at $Q_0^2$. 
As can be seen in Fig.~\ref{fig:GM_GE_mpi2} this linear behaviour
is well satisfied for 
 $Q_0^2=1.37$~GeV$^2$ the largest value shown in the figure.
Another option is to interpolate lattice data at $Q_0^2$ and perform
 a chiral extrapolation at this constant value of the momentum transfer
 squared.  Only the value $A$ of the slope should be
 affected whereas $f(Q_0^2,0)$ should not change. We checked these two 
procedures for various values
 of $Q^2$.
 We find that the results for $f(Q_0^2,0)$  
obtained  using these two procedures are indeed consistent.
  In what
 follows we will therefore use Eq.~(\ref{q2}) for the chiral extrapolation.

\begin{figure}[h]
\epsfxsize=8.truecm
\epsfysize=10truecm
\mbox{\epsfbox{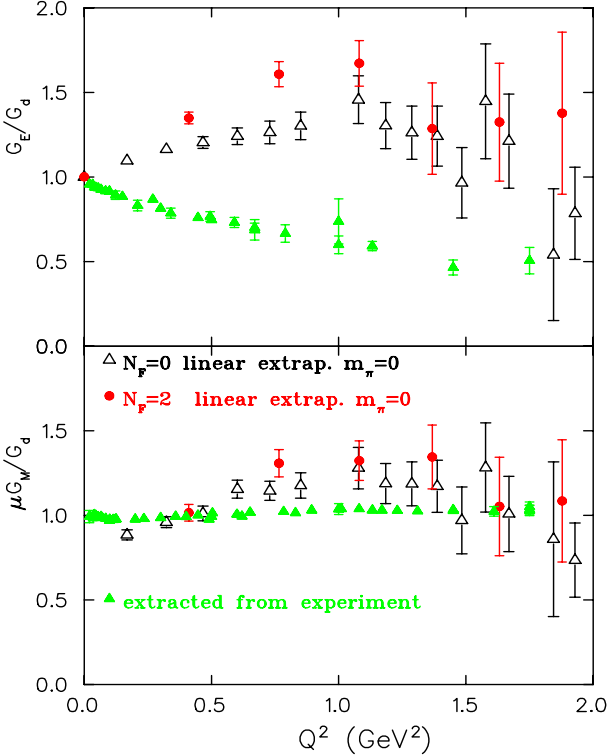}}
\caption{The ratios $G_E/G_d$ (upper) and  $\mu G_M/G_d$ (lower)
 as a function of  $Q^2$ at the chiral limit.  Open triangles show quenched results
obtained at the chiral limit
 by linear extrapolation of the form factors $G_M$ and $G_E$ and
filled circles denote the corresponding unquenched results.
The results
 for these isovector ratios extracted from experiment are shown by the filled
triangles.}
\label{fig:GM GE chiral}
\end{figure}

\begin{figure}[h]
\epsfxsize=8.truecm
\epsfysize=10truecm
\vspace*{0.5cm}
\mbox{\epsfbox{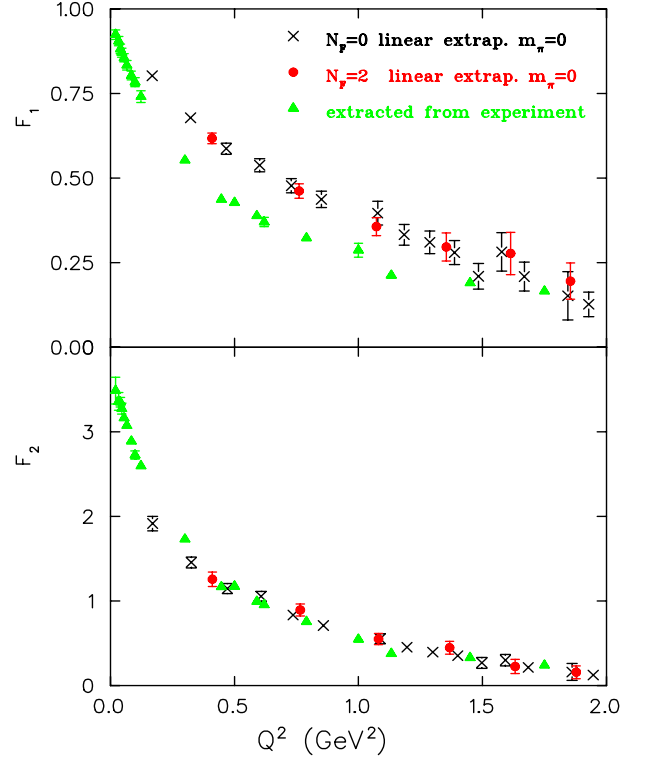}}
\caption{The form factors $F_1$  (upper) and $F_2$ (lower)
 as a function of  $Q^2$. Crosses show quenched results at the chiral limit,
and filled circles unquenched results.  The results
  extracted from experiment are shown by the filled
triangles.}
\label{fig:F1 F2 chiral}
\vspace*{-0.5cm}
\end{figure}

 The
resulting values at the chiral limit are shown in Fig.~\ref{fig:GM GE chiral}
for $G_E$ and $G_M$.
The disagreement with experiment
is larger in the case of the electric form factor and can be traced to its 
weak quark mass dependence. Lattice results also
show a different $Q^2$ dependence as compared to experiment.
The linearly extrapolated lattice results are closer to experiment in the case
of the magnetic form factor.
For comparison, we  show in Fig.~\ref{fig:F1 F2 chiral} 
 the lattice results for $F_1$ and $F_2$ after linearly extrapolated
to the chiral limit.
 There
is little deviation between  unquenched and quenched results
 at the chiral limit.
 In addition,  for $Q^2>0.5$~GeV$^2$ there is good agreement
between the lattice results for $F_2$ 
and the results extracted from experiment. This is not the case for $F_1$
where the experimentally determined isovector $F_1$ decays faster 
 as compared to the lattice results.  

\begin{figure}[h]
\vspace*{.3cm}
\epsfxsize=8.truecm
\epsfysize=10truecm
\mbox{\epsfbox{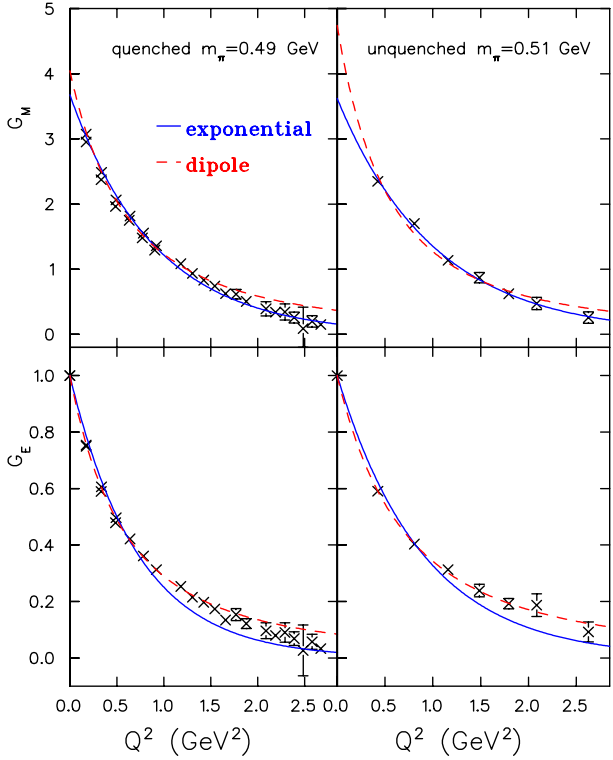}}
\caption{The magnetic (upper) and electric 
(lower) form factors for the quenched and the unquenched cases with fits to 
dipole (dashed line) and exponential (solid line) forms.}
\label{fig:dipole}
\end{figure}

In order to obtain the isovector magnetic moment $G_M(0)\def\kappa_{\rm v}$, 
one needs to evaluate the magnetic form factor $G_M$
at $Q^2 = 0$. This requires an extrapolation of lattice results 
to $Q^2=0$. One fitting Ansatz commonly used to describe the $Q^2$ dependence
of the form factors 
is a dipole. We thus consider a dipole form
 with different
isovector magnetic and electric dipole masses squared, $M_m$ and $M_e$:   
\beq
G_M(Q^2) &=& \frac{G_M(0)}{\left( 1 + \frac{Q^2}{M_m} \right)^2} \\
G_E(Q^2) &=& \frac{1}{\left( 1 + \frac{Q^2}{M_e} \right)^2} 
\eeq
A good fit 
giving $\chi^2/$(degree of freedom)~$\sim 1$ is obtained
when the form factors are fitted for $Q^2\stackrel{<}{\sim} 2.5$~GeV$^2$.
The quality of the fits is shown in Fig.~\ref{fig:dipole}, 
where we show quenched and unquenched data at $\kappa=0.1558$ and
$\kappa=0.1580$ respectively. On the same figure, we include
a fit to an exponential form. For the magnetic form factors, an exponential
{\it Ansatz} describes very well the $Q^2$ dependence, especially in the quenched 
case where it is in fact favored. An exponential fit, however, does not
provide a good fit to the electric form factors.
Therefore in order to extract the r.m.s. radii we will use throughout
a dipole {\it Ansatz}.
In Table~\ref{table:dipole} we give
 the magnetic moment, $G_M(0)$,
and the magnetic and electric dipole masses  
extracted
from the dipole fits to the quenched and unquenched results 
at each value of the quark mass. 
 The dipole masses extracted from the fits  are generally larger than the value
of  $M_d=0.71$~GeV$^2$ entering in the proton form factor  $G_d$. 
This is consistent with
the fact that the lattice data normalized with $G_d$
are not constant but increase as a function of $Q^2$ as shown in
 Figs.~\ref{fig:GEoverGd} and \ref{fig:GMoverGd}.  
In the same Table we also give the values extracted by applying dipole fits to
 the form factors after they have been 
linearly extrapolated to the chiral limit. Since the
dipole masses decrease with the quark mass, the fits of the form factors
at the chiral limit yield
a smaller value for $M_e$ and $M_m$. 
In the unquenched case these values are rather close to the value of the
 proton dipole mass, $M_d$.

\begin{widetext}
\begin{center}
\begin{table}[h]
\caption{
}
\label{table:dipole}
\begin{tabular}{|c|c|c|c|c|c|}
\hline
\multicolumn{1}{|c|}{$\kappa$ } &
\multicolumn{1}{ c|}{$G_M(0)$  } &
\multicolumn{1}{ c|}{$M_m$(GeV$^2$)} &
\multicolumn{1}{ c|}{ $M_e$(GeV$^2$)} & 
\multicolumn{1}{ c|}{ $<r_1^2>^{1/2}$(fm)} & 
\multicolumn{1}{ c|}{ $<r_2^2>^{1/2}$(fm)} 
\\
\hline
\multicolumn{6}{|c|}{Quenched $32^3\times 64$ \hspace*{0.5cm} $a^{-1}=2.14(6)$ GeV} \\
  0.1554 & 4.11(7) & 1.29(4) & 1.24(1) & 0.520(5) & 0.64(1)   \\
  0.1558 & 4.02(8) & 1.28(4) & 1.15(1) & 0.538(6) & 0.64(1) \\
  0.1562 & 3.90(9) & 1.19(4) & 1.08(1) & 0.550(8) & 0.66(1)  \\
 $\kappa_c=$0.1571   & 3.73(13)& 1.03(5) & 0.90(2) & 0.585(13) & 0.72(2)  \\
\hline
\multicolumn{6}{|c|}{Unquenched $24^3\times 40$  \hspace*{0.3cm} $a^{-1}=2.56(10)$~GeV} 
 \\
  0.1575  & 4.45(14)  & 1.53(7)  & 1.55(1) & 0.467(7) &0.58(2)      \\  
  0.1580  & 4.34(43)  & 1.23(16) & 1.41(2) & 0.462(23) & 0.67(5)  \\
\multicolumn{6}{|c|}{Unquenched $24^3\times 32$  \hspace*{0.3cm} $a^{-1}=2.56(10)$~GeV} \\
  0.15825 & 4.10(46)  & 1.17(17) & 1.19(4) & 0.500(29) & 0.68(6) \\
 $\kappa_c=$0.1585   & 3.25(48)& 0.792(17) &  0.66(4) & 0.756(36) & 0.79(13)  \\
\hline
\end{tabular}
\end{table} 
\end{center}
\end{widetext}

\begin{figure}[h]
\epsfxsize=8.truecm
\epsfysize=10truecm
\mbox{\epsfbox{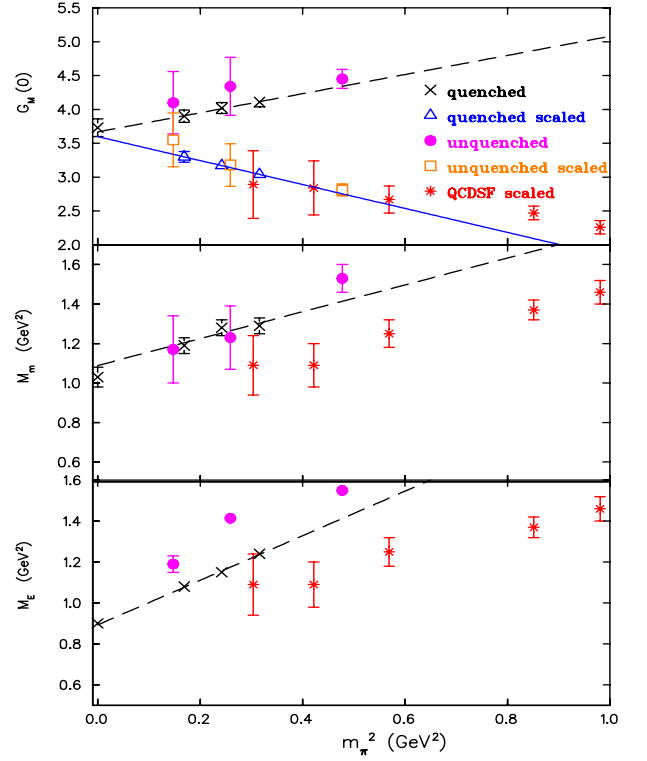}}
\caption{The magnetic moment (upper), the magnetic (middle) and electric 
(lower) dipole mass  extracted from fits, assuming a dipole dependence for the
form factors,are shown as a function of the pion mass squared. The dashed line is 
a linear fit to the quenched results.  The values at the chiral
limit are obtained from fitting the linearly extrapolated form factors
at the chiral limit to a dipole Ansatz. Our quenched data are
shown by the crosses, the unquenched data by the filled circles, 
our scaled quenched (unquenched) magnetic moments
by the open triangles (open triangles) and the quenched results at $\beta=6.0$ 
of Ref.~\cite{QCDSF} by the asterisks.} 
\label{fig:moments}
\end{figure}

\begin{figure}[h]
\epsfxsize=8.truecm
\epsfysize=10truecm
\vspace*{0.5cm}
\mbox{\epsfbox{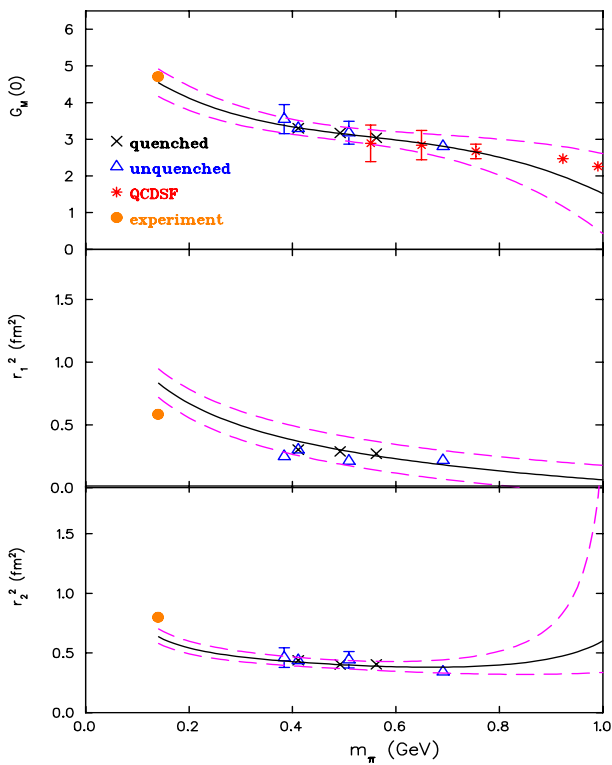}}
\caption{Chiral extrapolation of the magnetic moment (upper)  
and the r.m.s radii  $r_1$ (middle) and $r_2$ (lower). The solid line is 
the best fit to the effective chiral theory results. The dashed lines
show the maximal allowed error band using the errors on the fitted
parameters.}
\label{fig:chiral fits}
\vspace*{-0.5cm}
\end{figure}

\begin{table}[h]
\caption{The first column lists the fixed parameters and the second
their values at the physical pion mass. The third column gives the
fitted parameters and the fourth the values determined by
fitting to Eq.~(\ref{kappav}) for the magnetic moment, to Eq.~(\ref{rad1})
for $r_1^2$ and to Eq.~(\ref{rad2}) for $r_2^2$.} 
\label{table:chiral parameters}
\begin{tabular}{|c|c|c|c|}
\hline
\multicolumn{1}{|c|}{Fixed } &
\multicolumn{1}{ c|}{Empirical value } &
\multicolumn{1}{ c|}{Fitted } &
\multicolumn{1}{ c|}{Fitted value } 
\\
\multicolumn{1}{|c|}{parameter} &
\multicolumn{1}{ c|}{ } &
\multicolumn{1}{ c|}{parameter} &
\multicolumn{1}{ c|}{ } 
\\
\hline
$g_A$ & 1.267           & $\kappa_{\rm v}(0)$ & 6.08(11) \\
$c_A$ & 1.125           & $c_V $ & -2.75(50) GeV$^{-1}$ \\
$F_\pi$ & 0.0924~GeV    & $E_1$ & -5.60 (5) GeV$^{-3}$\\
$M_N$ & 0.9389~GeV      & $B_{10}$ & -0.3(3) GeV$^{-3}$ \\
$\Delta$ & 0.2711~GeV   & $B_{c2}$ & 0.61(4)\\
\hline
\end{tabular}
\end{table}

The quark mass dependence of the magnetic moments extracted
 from the dipole fits
is presented in Fig.~\ref{fig:moments}. A linear extrapolation 
of the quenched results to the chiral 
limit leads to $G_M(0) = 3.67(3) $. As expected, this value is
in agreement with the value obtained 
from the dipole fit of $G_M$ at  the chiral limit
 quoted in Table~\ref{table:dipole}.  This agreement confirms that our
extrapolation using Eq.~(\ref{q2}) yields results consistent to those
 obtained by extrapolating to the chiral limit at constant value of $Q^2$.
The slope of $F_1$ at $Q^2=0$ determines the transverse size of the hadron,
$<r_\perp^2> = -4 dF_1/dQ^2|_{Q^2=0}$. In
the non-relativistic limit  the r.m.s. radius is related to
the slope of the form factor at zero momentum transfer. Therefore 
the r.m.s. radii can
 be directly obtained  from the values of the dipole masses by using
\be
<r_i^2>=-\frac{6}{F_i(Q^2)}\frac{dF_i(Q^2)}{dQ^2}|_{Q^2=0}=\frac{12}{M_i} \hspace*{0.5cm} i=1,2 \quad.
\ee
The electric and magnetic radii are 
 given by $<r_{e,m}^2>=12/M_{e,m}$ and can be directly evaluated from
the values given in Table~\ref{table:dipole}.
We can also
 obtain $<r_i^2>$ in terms of $M_m$ and $M_e$  
using the relations
\be
<r_1^2>=\frac{12}{M_e}-\frac{3 F_2(0)}{2M_N^2} \hspace*{0.5cm}
<r_2^2>=\frac{12(1+F_2(0))}{F_2(0) M_m}-\frac{<r_1^2>}{F_2(0)} \quad.
\label{radii}
\ee
 In Eq.~(\ref{radii}) we 
take $F_2(0)= G_M(0)-1$ extracted from the dipole fits.
Alternatively, one can fit directly the $F_1$
and $F_2$ form factors and obtain the dipole masses $M_1$ and $M_2$. 
What one finds
via this procedure
is that the values for $<r_i^2>$ tend to be smaller than
but consistent within errors with the ones 
extracted using Eq.~(\ref{radii}). 
The quark mass dependence of the magnetic and electric dipole masses 
is also
 shown in Fig.~\ref{fig:moments}.
 A linear dependence in  $m_\pi^2$  
is consistent for the quenched results yielding, at the chiral limit,
 $M_m = 1.09(10)$~GeV$^2$ and $M_e = 0.89(4)$~GeV$^2$. 
Again, these values are in agreement with
the values extracted by fitting the form factors after they have
been linearly extrapolated to the chiral limit as can be seen from
the values quoted in Table~\ref{table:dipole}. 
After scaling the magnetic moments by the ratio of the physical nucleon
mass to  the one measured on the lattice, they
become an increasing faction of $m_\pi^2$ as can be seen 
in Fig.~\ref{fig:moments}. 
In the same figure we also  include
the quenched results obtained  from Ref.~\cite{QCDSF},
that used perturbatively improved Wilson fermions. We choose data at
$\beta=6.0$ where the value of the lattice spacing extracted from their
nucleon mass is $a^{-1}=1.83$ GeV. This is close enough to our quenched
lattice to allow a meaningful comparison.    
It is reassuring that, despite the fact that different currents
and Wilson fermions were used in the two calculations, results 
at a similar pion mass
 are consistent.

As the pion mass decreases, one expects cloud pion contributions to become
important and deviations from the linear dependence on $m_\pi^2$ should
be observed, in particular
at low $Q^2$, thus  affecting the values of $G_M(0)$   
and the dipole masses and hence the r.m.s. radii. 
In a recent calculation the quark mass 
dependence of the isovector magnetic moment and
radii was determined. This was done
within a
chiral effective theory with explicit nucleon and $\Delta$ degrees of 
freedom~\cite{chiral,QCDSF}. 
The isovector anomalous magnetic moment
as a function of the pion mass to one-loop order 
is given by~\cite{chiral}
\beq
\kappa_{\rm v}(m_\pi)&=&\kappa_{\rm v}(0) -\frac{g_A^2 m_\pi M_N}{4\pi F_\pi^2}
\nonumber \\ 
&+&\frac{2c_A^2\Delta M_N}{9\pi^2 F_\pi^2}\biggl [R_1(m_\pi) 
+ \log\left (\frac{m_\pi}{2\Delta}\right)\biggr] \nonumber \\ 
&-&8E_1 M_N m_\pi^2 + \frac{4c_A c_V g_A M_N m_\pi^2}{9\pi^2F_\pi^2}
\log\left(\frac{2\Delta}{\lambda}\right) 
\nonumber \\ 
 &+& \frac{4c_A c_V g_A M_N m_\pi^3}{27 \pi F_\pi^2 \Delta}
-\frac{8c_A c_V g_A \Delta^2 M_N}{27 \pi^2 F_{\pi}^2}  \biggl[ \nonumber \\
&\>&\hspace*{-1cm} \left(1-\frac{m_\pi^2}{\Delta^2}\right) R_1(m_\pi) 
+ \left(1-\frac{3m_\pi^2}{2\Delta^2}\right) \log\left(\frac{m_\pi}{2\Delta}
\right) \biggr ]
\label{kappav}
\eeq
where
\be
 R_1(m)=\frac{\sqrt{\Delta^2-m^2+i\epsilon}}{2\Delta} 
\log\left(\frac{\Delta+\sqrt{\Delta^2-m^2+i\epsilon}}{\Delta-\sqrt{\Delta^2-m^2+i\epsilon}}\right) 
\label{R1}
\ee 
and $\Delta=M_\Delta-M_N$ is the $\Delta$-nucleon mass splitting.
Following Ref.~\cite{chiral}
we fix $g_A$, $c_A$, $F_\pi$, $M_N$ and $\Delta$ to their physical
values given in Table~\ref{table:chiral parameters}
and vary $\kappa_{\rm v}(0)$, $c_V$ and $E_1$.
The counter term $E_1$ depends on the regularization scale $\lambda$ for which
we take  $\lambda=0.6$~GeV in order to 
make contact with Ref.~\cite{chiral}.
As can be seen in Fig.~\ref{fig:moments} the magnetic moments in the quenched
and unquenched theory are in agreement and therefore we use both
sets to fit to the chiral effective theory result given in Eq.~(\ref{kappav}).
Fitting to the rescaled data 
we obtain the curve  shown by the solid line 
in Fig.~\ref{fig:chiral fits}. 
In Table~\ref{table:chiral parameters} we give the values of 
$\kappa_{\rm v}(0)$, $c_V$ and $E_1$ 
extracted from the fit. The dashed lines give the maximal
 error band determined
by varying the
fitted parameters  by the quoted errors.
The extrapolated value of the magnetic moment at the physical pion mass is in agreement with 
experiment.

For the isovector Dirac and Pauli radii we use the one-loop results
given in Ref.~\cite{QCDSF}:
\beq
r_1^2 &=& -\frac{1}{(4\pi F_\pi)^2}\biggl[1+7g_A^2+\left(10g_A^2+2\right)
\log\left(\frac{m_\pi}{\lambda}\right)\biggr] \nonumber \\
&\>&-\frac{12 B_{10}}{(4\pi F_\pi)^2} + \frac{c_A^2}{54\pi^2 F^2_\pi}
\biggl[26 + 30 \log\left(\frac{m_\pi}{\lambda}\right) \nonumber \\
&\>&+30 R_2(m_\pi) \biggr]
\label{rad1}
\eeq
and
\beq
r_2^2&=& \frac{1}{\kappa_{\rm v}(m_\pi)}\biggl\{ 
\frac{g_A^2 M_N}{8F_\pi^2\pi m_\pi} 
+ \frac{c_A^2 M_N}{9F^2_\pi \pi^2\Delta} R_2(m_\pi) \nonumber \\
&\>& +24 M_N B_{c2} \biggr\} 
\label{rad2}
\eeq
where
\be
 R_2(m)=\frac{\Delta}{2\sqrt{\Delta^2-m^2+i\epsilon}} 
\log\left(\frac{\Delta+\sqrt{\Delta^2-m^2+i\epsilon}}{\Delta-\sqrt{\Delta^2-m^2+i\epsilon}} \right)\quad.
\label{R2}
\ee 
The only parameter that we vary in fitting the Dirac radius is the counter term 
$B_{10}$, which depends on the scale $\lambda$ and parametrizes short distance
 contributions. Once the magnetic moment is fitted, the only parameter entering
in the  Pauli radius that we vary 
is the counter term $B_{c2}$, which is the analogue of
 $B_{10}$. The resulting fits for the radii are shown 
in Fig.~\ref{fig:chiral fits}. The pion mass dependence of the  Dirac radius  
is not well reproduced. Since this is related to the slope of $F_1$ this
is not surprising given that the lattice results have a different slope from 
the experimental one and hardly show any quark mass dependence.

\section{Conclusions}

The elastic isovector 
nucleon form factors are calculated in lattice QCD with Wilson
fermions both in the quenched approximation and using 
unquenched configurations~\cite{newSESAM,Carsten}
 with two flavors of dynamical Wilson fermions. The current
work presents an improvement to
a previous lattice study~\cite{QCDSF}, carried out in the
quenched approximation, in a number of ways: In the quenched theory 
we use a lattice
of twice the spatial and temporal size. This allows an accurate determination
of the form factors at lower values of $Q^2$, enabling us to extract more
reliably the dependence on $Q^2$. 
In addition, 
the quenched calculation
is carried out at smaller quark masses, bringing us closer
to the chiral limit.  Preliminary quenched results on
a lattice
of size $32^3\times 48$ at $\beta=6.0$ using Wilson fermions
were presented in Ref.~\cite{QCDSF2}. Although the low $Q^2$
range probed is the same as in the current work, only
results after linear extrapolation 
to the chiral limit, using data computed at two 
light quark masses and thus carrying large statistical errors, 
were discussed.
 Furthermore in this work we evaluated the form factors
  in the unquenched theory allowing us
to assess unquenching effects
for pion masses down to about 380 MeV. Finally, an 
improved overconstrained analysis is carried out
where the nucleon source is optimized and all the lattice momentum vectors
 contributing to a given value of $Q^2$ are taken into account.
The
 resulting statistical errors are therefore small enough that
 a comparison between quenched and unquenched results is meaningful.
 What we find is that both quenched and unquenched results for
 both form factors decrease with the
 quark mass. Unquenching effects are
 small and
 the results obtained after
 a linear extrapolation in $m_\pi^2$ to the chiral limit
 fall on the same curve
 as can be seen for example
 in Fig.~\ref{fig:F1 F2 chiral} for $F_1$ and $F_2$.
Assuming lattice artifacts are under control, 
this improved analysis gives results that can  be compared to experiment
 by extracting the isovector form factors
from the proton and neutron measurements of these quantities. 
The largest uncertainties regarding our lattice results 
are how close to the continuum limit these
results are and the chiral extrapolation.
With these caveats in mind, the
comparison of the
results obtained here to experiment
reveals interesting features: Both quenched and  unquenched
results are higher than the experimentally extracted form factors, with
the deviations being larger in the case of the electric isovector form
factor. In the quenched case, where we have very accurate
results at low momentum transfer, we find that 
 the electric form factor decreases slower with $Q^2$
compared to what is observed  experimentally.
This different behavior is also reflected 
in the ratio $\mu G_E/G_M$, where the lattice
results are constant up to about $Q^2=2.5$~GeV$^2$ whereas the experimentally
determined data decrease as a function of $Q^2$. This leads to
  smaller
values for the r.m.s.
 radii showing that pion cloud contributions are smaller at these
quark masses than in the physical regime. In the
range of quark masses investigated in this work,
the quark mass dependence observed for  $\mu G_E/G_M$ is 
small and so are unquenching effects.
Using  chiral effective theory to one-loop to extrapolate
the magnetic moment to the chiral limit, we find that the
lattice results extrapolate nicely to the experimental value.
The charge radius, 
on the other hand, is constant over the range of quark masses used in
this work and therefore deviates from experiment. 
This is again related to the deviation observed between experiment and
lattice results in the case of the charge form factor.
Since the lattice size in the quenched case is large, we do not expect finite
volume effects to be the reason for the discrepancy.
 What needs to be checked is finite lattice spacing effects and 
whether we are close enough to the chiral limit.
The observed disagreement with experiment in the case of
 the charge form factor is puzzling and
a study using finer lattices should follow.
If one requires in addition
 dynamical fermions and small quark masses to be closer to the chiral limit, 
then 
such a study would require large computer resources.
Unquenched configurations with pion masses down 
to  250 MeV on reasonably large and fine lattices will become
available in the near future enabling us, using the techniques
of the current work, to obtain results closer to the physical regime
avoiding uncontrolled extrapolations.

{\bf Acknowledgments:} 
We would like to thank B. Orth,  Th. Lippert and K. Schilling~\cite{newSESAM}
as well as  C. Urbach, K. Jansen, A. Shindler 
and U. Wenger~\cite{Carsten} 
 for providing the unquenched
configurations used in this work.
A.T. would like to acknowledge support by the University of Cyprus and the program
``Pythagoras'' of the Greek Ministry of Education.
C.A. and J.W.N. would like to thank the Institute for Nuclear Physics at the University of Washington for its hospitality and
 partial support during the completion of this work.
The computations for this work were partly 
carried out on the IBM machine at NIC, Julich,
Germany. This work is 
supported in part by the  EU Integrated Infrastructure Initiative
Hadron Physics (I3HP) under contract RII3-CT-2004-506078 and by the 
U.S. Department of Energy (D.O.E.) Office of Nuclear Physics under contract DF-FC02-94ER40818
This research used resources of the National Energy Research Scientific 
Computing Center, which is supported by the Office of Science of the U.S. 
Department of Energy under Contract No. DE-AC03-76SF00098.

\end{document}